\documentclass[12pt]{aastex}
\usepackage[twocolumn]{emulateapj5}

\shorttitle{Radiative acceleration and transient, radiation-induced
electric fields}
\shortauthors{Zampieri et al.}

\begin{document}

\title{Radiative acceleration and transient, radiation-induced \\
electric fields}
\author{L. Zampieri}
\affil{INAF-Osservatorio Astronomico di Padova, Vicolo dell'Osservatorio
5, I-35122 Padova, Italy}

\author{R. Turolla}
\affil{Dipartimento di Fisica, Universit\`a di Padova, Via Marzolo 8,
I-35131, Padova, Italy}

\author{L. Foschini}
\affil{IASF-CNR, Sezione di Bologna, Via Gobetti
101, I-40129 Bologna, Italy}

\and

\author{A. Treves}
\affil{Dipartimento di Scienze, Universit\`a dell'Insubria,
Via Valleggio 11, I-22100 Como, Italy}

\begin{abstract}
The radiative acceleration of particles and the electrostatic
potential fields that arise in low density plasmas hit by radiation
produced by a transient, compact source are investigated. We calculate
the dynamical evolution and asymptotic energy of the charged particles
accelerated by the photons and the radiation-induced electric double
layer in the full relativistic, Klein-Nishina regime. For fluxes in
excess of $10^{27}$ ${\rm erg}\, {\rm cm}^{-2}\, {\rm s}^{-1}$, the
radiative force on a diluted plasma ($n\la 10^{11}$ cm$^{-3}$) is so
strong that electrons are accelerated rapidly to relativistic speeds
while ions lag behind owing to their larger inertia. The ions are
later effectively accelerated by the strong radiation-induced double
layer electric field up to Lorentz factors $\approx 100$, attainable
in the case of negligible Compton drag. The asymptotic energies
achieved by both ions and electrons are larger by a factor 2--4 with
respect to what one could naively expect assuming that the
electron-ion assembly is a rigidly coupled system. The regime we
investigate may be relevant within the framework of giant flares from
soft gamma-repeaters.
\end{abstract}


\section{Introduction}

Particle acceleration is an ubiquitous feature in many classes of
astrophysical objects, such as active galactic nuclei, gamma-ray
bursts, pulsars. Despite several physical processes are know to be
responsible for accelerating particles under different conditions,
radiative acceleration has special importance. \citet{GUREVICH}
were the first to study the acceleration of charged particles
caused by scattering off a radiation beam. They derived the
relativistic equation of motion for the electrons and assumed that the
radiative force is applied directly to the electron-ion assembly. For
the typical radiative fluxes produced in Type II supernovae, they
found that ions may acquire energies of the order of their rest mass
energy.  Radiative acceleration was the subject of several further
investigations in the '80s, mainly connected with the possibility that
this mechanism could drive relativistic collimated outflows, or jets,
such as those observed in Active Galactic Nuclei (AGNs) and in some
Galactic sources (e.g. SS433).  Earlier studies (e.g. \citealt{ABRA};
\citealt{ODELL}; \citealt{ODELLb}; \citealt{CHENG}; \citealt{SIKORA};
\citealt{PHINNEY}) dealt with the radiative acceleration of electrons
assuming Thomson scattering in the particle frame and adopting a
test-particle approach. This is fully justified when the medium is
optically thin (for a discussion of the optically thick case see
\citealt{TUROLLA1}) and allowed to clarify the role of
Compton drag in establishing the limiting plasma velocity and the
dynamical effects caused by anisotropic Compton losses of plasma
internal energy (``Compton rocket'').

The first calculation of the radiative force acting on electrons
taking into account the Klein-Nishina cross section was performed by
\citet{BLUM}. More recently, \citet{MADAU}
re-addressed the issue of the dynamics of a cold plasma subject to an
external radiation field in the Klein-Nishina regime. In a companion
paper \citep{THOMPSON}, the feedback of pair creation on
radiative acceleration was also considered. Contrary to earlier works,
the main focus of these investigations is on the dynamics of the
fireball expected to energize gamma-ray bursts. In such a scenario,
the mean photon energy is large enough to give raise to copious pair
production.

An interesting and related research thread is the acceleration by
electrostatic fields generated by space charge
separation. \citet{SHVARTSMAN} considered a mechanism suitable for
positron acceleration in accreting neutron stars. He proposed that the
radiation field incident on the plasma acts differently on electrons
and ions because of the large difference in mass and cross section
(cf. also \citealt{MARASCHI1}; \citealt{MARASCHI2}). Electrons are separated from
ions and an electrostatic field arises to restore charge
neutrality. The net electrostatic potential is able to accelerate
positrons, which are limited by pair production and inverse Compton,
but reach energies of tens of MeV. The calculation has been revised by
\citet{TUROLLA2}, who found that the maximum energy attainable is
limited to $\sim$ 400 keV. \citet{TREVES1} and \citet{TREVES2}
investigated this scenario in connection with the vacuum
breakdown near an accreting black hole, while \citet{HIROTANI1}
studied the case of a supermassive black hole in an AGN.
\citet{LITWIN} proposed another version, consisting of
infalling blobs of plasma onto a neutron star, in which the electric
field is generated by the co-rotation of the strong magnetic field of
the star (cfr. also \citealt{HIROTANI2}). This mechanism has also
several applications to particle acceleration in solar flares and
planetary bow shocks (cf. \citealt{BALL} and references therein).

In this Paper, we present a complete analysis of the acceleration of
electrons and ions by electrostatic potential fields that arise in a
low density plasma illuminated by a burst of radiation. Extending
previous works (in particular that by \citealt{MADAU}), the
dynamical evolution of the charged particles under the joint action of
the internal electrostatic and the external radiation forces is
investigated. In \S~2 we derive the non-relativistic and relativistic
equations of motion for the charged plasma particles in presence of a
Double Layer. Our results are presented in \S 3 and thoroughly
discussed in \S 4. Conclusions follow in \S 5.

\section{Radiative acceleration and electric restoring force in a plasma}

In a plasma, any slight deviation from a condition of global electric
neutrality gives rise to large restoring forces. Radiation impinging
on a plasma acts differently on ions and electrons, because of their
large difference in mass, generating a space charge
separation. Therefore, an electric field rises to restore global
neutrality. Since the resulting
electric field is across two opposite space charge layers, it is often
called ``double layer'' (DL), particularly with reference to studies
in interplanetary bow shocks and solar flares (see e.g. \citealt{BLOCK};
\citealt{RAADU}).

Consider a transient, compact source of high energy photons of
luminosity $L$ surrounded by an optically thin shell of cold
plasma. The plasma is composed of electrons and completely ionized
ions of atomic number $Z$. The radiation flux is sufficiently strong
that gravity can be neglected and, at the typical distances from the
central source considered here, magnetic fields are assumed to be
dynamically unimportant. The equations of motion of the electrons and
ions (in a reference frame ${\bf K}$ at rest with respect to the
central source) under the joint action of the radiative force and the
restoring force of the DL are
\begin{eqnarray}
&& \gamma_e \frac{d}{dt} \left(\gamma_e m_e {\bf v}_e \right) =
{\bf F}_{E,e} + {\bf F}_{rad} \\
&& \gamma_i \frac{d}{dt} \left(\gamma_i m_i {\bf v}_i \right) =
{\bf F}_{E,i} \, ,
\label{e:rel}
\end{eqnarray}
where $\gamma_{e,i}$, ${\bf v}_{e,i}$ and $m_{e,i}$ are the Lorentz
factor, 3-velocity and mass of electrons and ions, and ${\bf
F}_{rad}$, ${\bf F}_{E,e}$ and ${\bf F}_{E,i}$ are the spatial
components of the radiative and electric 4-forces acting on electrons
and ions. The radiative force acting on the ions is neglected because
it is a factor $(m_e/m_i)^2$ smaller than that on the electrons. Here
we consider mono-energetic particles. If the charged particles have a
momentum distribution, the previous equations must be averaged over
the normalized phase-space density $f_{e,i}({\bf v}_{e,i})$.  We
checked a posteriori that the radiation-reaction force due to the
energy radiated by the accelerated particles is negligible and does
not influence the dynamics.

The electric force is given by the spatial components of the Lorentz
4-force vector
\begin{eqnarray}
&& {\bf F}_{E,e} = - \gamma_e Z e {\bf E} \\
&& {\bf F}_{E,i} =   \gamma_i Z e {\bf E} \, ,
\end{eqnarray}
where ${\bf E}$ is the electric field and $Z$ the atomic number of the
ions. The general expression for the spatial part of the radiative
4-force acting on the electrons is (cfr. \citealt{BLUM}; \citealt{MADAU})
\begin{equation}
{\bf F}_{rad} = - \int \Delta \left(\frac{\epsilon}{c} \, {\bf k}\right)
\gamma_e \left( 1-\frac{{\bf v}_e\cdot{\bf k}}{c} \right)
\frac{I_{\epsilon}}{\epsilon}
\frac{d\sigma}{d\Omega_s} d\epsilon d\Omega d\Omega_s \, ,
\end{equation}
where $\epsilon$ and ${\bf k}$ are the photon energy and unit
propagation vector, $I_\epsilon$ is the radiation specific
intensity, $d\sigma/d\Omega_s$ is the differential scattering
cross section, $\Omega$ is the solid angle subtended by the
source and $\Omega_s$ that over the scattering angle.
The first term
$\Delta (\epsilon {\bf k}/c) = (\epsilon_1 {\bf k}_1 - \epsilon {\bf k})/c$
is the photon momentum change in a scattering ($\epsilon_1$ and
${\bf k}_1$ are the photon energy and unit propagation vector after
scattering), while the second
$\gamma_e [ 1-({\bf v}_e\cdot{\bf k}/c] (I_{\epsilon}/{\epsilon}) (d\sigma/d\Omega_s) d\epsilon d\Omega d\Omega_s$
represents the rate at which photons with energies between $\epsilon$ and
$\epsilon+d\epsilon$ and directions between $\Omega$ and
$\Omega+d\Omega$ are Compton scattered by electrons with scattering
angle between $\Omega_s$ and $\Omega_s+d\Omega_s$.
The general expression for the differential scattering cross section
is given by the Klein-Nishina formula
\begin{equation}
\frac{d\sigma}{d\Omega_s^{'}} = \frac{3}{16\pi} \sigma_T
\left(\frac{\epsilon_1^{'}}{\epsilon^{'}}\right)^2
\left(\frac{\epsilon^{'}}{\epsilon_1^{'}} + \frac{\epsilon_1^{'}}{\epsilon^{'}}
- \sin^2\chi_s^{'} \right) \, ,
\label{e:kn}
\end{equation}
where $\sigma_{T}=6.65\times 10^{-25}$ cm$^{-2}$ is the Thomson cross
section, primed quantities are evaluated in the reference frame ${\bf
K}^{'}$ comoving with the electron and $\chi_s^{'}$ is the (polar) angle
between the incident and scattered photon
directions. Equation~(\ref{e:kn}) must be supplemented with the
Compton formula for the energy of the scattered photon
\begin{equation}
\epsilon_1^{'} = \frac{\epsilon^{'}}{1+(\epsilon^{'}/m_ec^2)(1-\cos\chi_s^{'})}
\, .
\label{e:compt}
\end{equation}
The corresponding unprimed quantities in the system frame ${\bf K}$
are obtained Lorentz transforming photon energies and directions by
means of the the Doppler and aberration formulas.

In the following we will investigate radiative acceleration from an
isotropic, spherically symmetric source of radiation. Then, the motion
of the charged particles is radial and the equations simplify to
\begin{eqnarray}
&& \gamma_e \frac{d}{dt} \left(\gamma_e m_e v_e \right) = F_{E,e} + F_{rad}
\label{e:eqrad1} \\
&& \gamma_i \frac{d}{dt} \left(\gamma_i m_i v_i \right) = F_{E,i} \, ,
\label{e:eqrad2}
\end{eqnarray}
where $v_{e,i}$ are the radial components of the electron and ion
velocities, and $F_{rad}$, $F_{E,e}$ and $F_{E,i}$ are the radial
components of the radiative and electric 4-forces acting on electrons
and ions. They are given by
\begin{eqnarray}
&& F_{E,e} = - \gamma_e Ze E \\
&& F_{E,i} =   \gamma_i Ze E \\
&& F_{rad} = - \int \frac{\epsilon_1 \mu_1 - \epsilon \mu}{c}
\gamma_e \left( 1-\frac{v_e}{c} \mu \right) \frac{I_{\epsilon}}{\epsilon}
\frac{d\sigma}{d\Omega_s} d\epsilon d\Omega d\Omega_s \, ,
\label{e:frad}
\end{eqnarray}
where $E$ is the radial component of the electric field, and
$\mu$ ($\mu_1$) is the cosine of the angle between the radial and
incident (scattered) photon directions. In terms of these angles and the
polar $\chi_s^{'}$ and azimuthal $\phi_s^{'}$ scattering angles in
${\bf K}^{'}$, it is
$\mu_1^{'} = \mu^{'} \cos\chi_s^{'} + (1-{\mu^{'}}^2)^{1/2} \sin\chi_s^{'} \cos\phi_s^{'}$.

If there are no external magnetic fields and the plasma particles move
along the radial direction, the radial component of the electric field
$E$ can be obtained from the Gauss theorem. For a (geometrically thin)
DL, we have
\begin{equation}
E = 4\pi Ze n \Delta x \, ,
\label{e:campoel}
\end{equation}
where $n$ is the volume density of charged particles and
$\Delta x=r_e-r_i\ll r_{e,i}$ is the displacement between the electron and ion
positions $r_e$ and $r_i$. We checked a posteriori that the condition
$\Delta x \ll r_{e,i}$ is satisfied.

\subsection{The non-relativistic equations in the Thomson regime}

The Newtonian equations for the radial motion of the electrons and
ions under the joint action of the radiative force and the restoring
force of the DL can be obtained from
equations~(\ref{e:eqrad1})--(\ref{e:frad}) neglecting terms
$O(v/c)$. They read
\begin{eqnarray}
&& m_e\frac{d^2x_e}{dt^2} = -Ze E + F_{rad,T}^{nr} \label{e:nonrela} \\
&& m_i\frac{d^2x_i}{dt^2} = Ze E \, ,
\label{e:nonrelb}
\end{eqnarray}
where $x_{e,i}$ are the displacements of electrons and ions from their
initial position $R$ ($x_{e,i}=r_{e,i}-R$). For Thomson scattering
($\epsilon_1^{'}\simeq\epsilon^{'}$) equation~(\ref{e:frad}) gives
\begin{equation}
F_{rad,T}^{nr}=\sigma_{T} \frac{F}{c} \, ,
\label{e:eradt}
\end{equation}
where $F = \int I_\epsilon \mu d\epsilon d\Omega = L/ 4 \pi r_e^2$ is
the radiative flux emitted by the central source at distance
$r_e=R+x_e$.

By substituting equations~(\ref{e:campoel}) and~(\ref{e:eradt}) into
equations~(\ref{e:nonrela}) and~(\ref{e:nonrelb}), we obtain
\begin{eqnarray}
&& \frac{d^2x_e}{dt^2}=-\frac{4\pi Z^2 e^{2} n}{m_e} \Delta x +
\frac{\sigma_{T} F}{m_e c} \label{e:nrel1} \\
&& \frac{d^2x_i}{dt^2}=\frac{4\pi Z^2 e^{2} n}{m_i} \Delta x \, .
\label{e:nrel2}
\end{eqnarray}
These are the equations of motion of a particle oscillating with the
plasma frequency (pulsation)
\begin{equation}
\omega_{e,i}=\sqrt{\frac{4\pi Z^2 e^{2} n}{m_{e,i}}} \, .
\label{e:pulse}
\end{equation}
In particular, the electron plasma frequency  is
\begin{equation}
\omega_e = 5.6\times 10^4 Z \left(\frac{n}{1 \ {\rm cm}^{-3}}\right)^{1/2}
{\rm s}^{-1} \, .
\label{e:epulse}
\end{equation}
In a fully ionized hydrogen plasma (electrons and protons only), $Z=1$
and $m_i=m_p$. Finally, we can write two coupled equations for the
displacements $x_e$ and $x_i$ in the form
\begin{eqnarray}
&& {\ddot x_e} = -\Delta x + \frac{\cal L}{(1+x_e/R)^2} \label{e:nonrel1} \\
&& {\ddot x_i} = \frac{m_e}{m_i} \Delta x \label{e:nonrel2} \, ,
\end{eqnarray}
where $\ddot x_{e,i} = d^2x_{e,i}/d(\omega_e t)^2$ and
\begin{eqnarray}
&& {\cal L} = \frac{\sigma_{T} L}{4 \pi R^2 c m_e \omega_e^2} \nonumber \\
&& = 62 \ Z^{-2} \left(\frac{n}{1 \ {\rm cm}^{-3}}\right)^{-1}
\left(\frac{R}{10^9 \ {\rm cm}}\right)^{-2} \times \nonumber \\
&& \qquad \qquad \quad
\left(\frac{L}{10^{38} \ {\rm erg s}^{-1}}\right) \ {\rm cm}
\label{e:lum}
\end{eqnarray}
represents the amplitude of the radiation-induced oscillations (for
small displacements $x_e\ll R$; see eq.~[\ref{e:nonrel1}]).

\subsection{The relativistic, Klein-Nishina regime}

The coupled radial equations for the motion of electrons and ions in
the reference frame ${\bf K}$ are (eqs.~[\ref{e:eqrad1}]
and~[\ref{e:eqrad2}])
\begin{eqnarray}
&& \gamma_e \frac{d}{dt} \left(\gamma_e m_e v_e \right) = - \gamma_e Ze E +
F_{rad} \label{e:rela} \\
&& \gamma_i \frac{d}{dt} \left(\gamma_i m_i v_i \right) =  \gamma_i Ze E \, .
\label{e:relb}
\end{eqnarray}
The radial component of the radiative force is calculated inserting
the Klein-Nishina and Compton formulas (eqs.~[\ref{e:kn}]
and~[\ref{e:compt}]) into equation~(\ref{e:frad}) and performing the
integral over $\Omega_s$. We obtain (see e.g. \citealt{BLUM}; \citealt{MADAU})
\begin{eqnarray}
F_{rad} &=& \frac{\sigma_T}{c} \gamma_e \int \frac{K(x^{'})}{x^{'}}
\left\{ \gamma_e^2 \left[ \mu (1+\beta_e^2) - \beta_e(1+\mu^2) \right]
\right. \nonumber \\
&+& \left. x^{'} \mu ( 1-\mu \beta_e) \right\} I_\epsilon d\epsilon d\Omega
\, ,
\label{e:fradrel}
\end{eqnarray}
where $\beta_e=v_e/c$,
$x^{'}=\gamma_e(\epsilon/m_ec^2)(1-\beta_e\mu)$ and
\begin{eqnarray}
K(x^{'}) &=& \frac{1}{\sigma_T} \int x_1^{'} (1-\cos\chi_s^{'})
\frac{d\sigma}{d\Omega_s^{'}} d\Omega_s^{'} \nonumber \\
&=& \frac{3}{4{x^{'}}^2} \left[ \frac{{x^{'}}^2 - 2x^{'} -3}{2x^{'}}
\ln(1+2x^{'}) \right. \nonumber \\
&+& \left. \frac{-10{x^{'}}^4 + 51{x^{'}}^3 + 93{x^{'}}^2 + 51 x^{'} + 9}
{3(1+2x^{'})^3} \right] \, .
\end{eqnarray}
For Thomson scattering, equation~(\ref{e:fradrel}) reduces to
\begin{equation}
F_{rad,T} = \frac{\sigma_T}{c} \gamma_e^3 \left[ (1+\beta_e^2)F - v_e (U+P)
\right] \, ,
\label{e:fradnr}
\end{equation}
where $U=c^{-1} \int I_\epsilon d\epsilon d\Omega$ and $P= c^{-1} \int
I_\epsilon \mu^2 d\epsilon d\Omega$ are the radiation energy density
and pressure. The same expression can be obtained Lorentz transforming
the radiative force from the reference frame comoving with the
electrons ${\bf K}^{'}$. In that frame it is $F_{rad}^{'} =
\sigma_T F^{'}/c$ and $F^{'}$ is the radiative flux measured in the
comoving frame. If the radiation field is stationary, the time
component of the radiative 4-force is zero. Thus, in ${\bf K}$ it is
$F_{rad} = \gamma_e F_{rad}^{'} = \gamma_e \sigma_T F^{'}/c$, where
$F^{'}$ is given by the Lorentz transformation for the radiative flux
\begin{equation}
F^{'} = \gamma_e^2 \left[ (1+\beta_e^2)F - v_e (U+P) \right] \, ,
\label{e:flusso}
\end{equation}
and one finally obtains equation~(\ref{e:fradnr}).

For a point-like source or sufficiently far from a finite-size source
the radiation
field is in radial streaming, i.e. $I_\epsilon = I_{\epsilon, 0} \,
\delta(\mu-1)$ and $F=cU=cP$ (in both ${\bf K}$ and ${\bf K^{'}}$). In
this limit equation~(\ref{e:fradnr}) reduces to
\begin{equation}
F_{rad,T}^s = \frac{\sigma_T F}{c} \gamma_e^3 (1-\beta_e)^2 \, .
\end{equation}
If, in addition, the input radiation spectrum is monochromatic,
$I_{\epsilon, 0} = I_0 \delta(\epsilon-\epsilon_0)$, then
equation~(\ref{e:fradrel}) becomes\footnote{Equation~(\ref{e:fradrels})
corrects a typo contained in a similar equation (eq.[48]) derived by
\citet{MADAU}.}
\begin{equation}
F_{rad}^s = \frac{\sigma_T F}{c} \gamma_e^3 (1-\beta_e)^2
\left[ \frac{K(x_0^{'})}{x_0^{'}}
\left(1+\frac{\epsilon_0}{\gamma_e m_e c^2}\right) \right] \, ,
\label{e:fradrels}
\end{equation}
where $x_0^{'} =
\gamma_e(\epsilon_0/m_ec^2)(1-\beta_e)$. The term in square brackets
represents the Klein-Nishina correction to the radiative
force. Finally, substituting equation~(\ref{e:fradrels}) into
equation~(\ref{e:rela}) and using the expression for the electric
field (eq.~[\ref{e:campoel}]), the equations of motion take the form
\begin{eqnarray}
&& \frac{d}{dt} \left(\gamma_e m_e v_e \right) = - 4 \pi Z^2 e^2 n \Delta x
\nonumber \label{e:rel1} \\
&& + \frac{\sigma_T F}{c} \,
\gamma_e^2 (1-\beta_e)^2 \left[ \frac{K(x_0^{'})}{x_0^{'}}
\left(1+\frac{\epsilon_0}{\gamma_e m_e c^2}\right) \right] \\
&& \frac{d}{dt} \left(\gamma_i m_i v_i \right) = 4 \pi Z^2 e^2 n \Delta x \, .
\label{e:rel2}
\end{eqnarray}
Introducing the new variables $y_{e,i}=v_{e,i}/\omega_e$,
equations~(\ref{e:rel1}) and~(\ref{e:rel2}) can be written
\begin{eqnarray}
&& \gamma_e^3 {\dot y_e} = - \Delta x + \gamma_e^2
\left(1 - \frac{\omega_e y_e}{c}\right)^2 \frac{\cal L}{(1+x_e/R)^2}
\times \nonumber \\
&& \qquad\qquad\qquad\qquad \left[ \frac{K(x_0^{'})}{x_0^{'}}
\left(1+\frac{\epsilon_0}{\gamma_e m_e c^2}\right) \right] \label{e:rel3} \\
&& \gamma_i^3 {\dot y_i} = \frac{m_e}{m_i} \Delta x \, ,
\label{e:rel4}
\end{eqnarray}
where $\dot y_{e,i} = dy_{e,i}/d(\omega_e t)$.

\section{Results}

Assuming that $x_{e,i} \ll R$ and $F=const$ (i.e. the luminosity of
the central source does not vary with time), the non-relativistic
equations (eq.~[\ref{e:nonrel1}] and~[\ref{e:nonrel2}]) can be solved
analytically. Taking as initial conditions $x_{e,i}(0) = 0$ and
$dx_{e,i}/dt|_0 = 0$, the displacements of electrons and protons are
given by
\begin{eqnarray}
&& x_e(t) = x_{eq} \left[ \frac{1}{2} (\omega_r^2 - \omega_e^2) t^2 +
\left(\frac{\omega_e}{\omega_r}\right)^2 \left(1 + \Theta \right)\right]
\label{e:sol1} \\
&& x_i(t) = x_{eq} \left[ \frac{1}{2} \omega_i^2 t^2 -
\left(\frac{\omega_i}{\omega_r}\right)^2 \left(1 + \Theta \right)\right] \, ,
\label{e:sol2}
\end{eqnarray}
where
\begin{eqnarray}
&& \Theta = \sin(\omega_r t - \pi/2) \\
&& \omega_r = \left[4\pi Z^2 e^2 n \left(\frac{m_e+m_i}{m_e m_i}\right)
\right]^{1/2} = \omega_e \left(1+\frac{m_e}{m_i}\right)^{1/2} \\
&& x_{eq} = \frac{\sigma_T F}{c m_e \omega_r^2} = {\cal L}
\left(1+\frac{m_e}{m_i}\right)^{-1} \, .
\end{eqnarray}
The relativistic equations (eqs.~[\ref{e:rel3}] and~[\ref{e:rel4}])
have been integrated numerically by using a $4^{th}$--order
Runge--Kutta method (see e.g. \citealt{NR}). The initial
conditions are the same as in the non-relativistic case. Once a
solution for $x_{e,i}$ is computed, the electric field of the DL is
calculated from equation~(\ref{e:campoel}). We performed the
integration for several values of the parameter ${\cal L}$ and for
flashes with two different durations ($\tau = 15.9 (2\pi/\omega_e)$ s
and $\tau = 4.8 \times 10^4 (2\pi/\omega_e)$ s). The results of our
calculations are shown in Figs.~\ref{fig3}--\ref{fig9} for $R=10^9$
cm, $n=1$ cm$^{-3}$, $Z=1$.

\section{Discussion}

Equations~(\ref{e:sol1}) and~(\ref{e:sol2}) show the properties of the
solution in the non-relativistic regime. The system is a coupled
oscillator with a driving force (radiation). When the radiation flash
hits the plasma shell, electrons are pushed outwards and begin to
oscillate. Because of the restoring force of the DL, ions are dragged
by the electrons. The whole system oscillates essentially at the
electron plasma frequency ($\omega_r/2\pi \simeq \omega_e/2\pi$) and
is accelerated outwards.  The amplitude of the oscillations of the
ions is a factor $m_e/m_i$ smaller than that of the electrons. From
equation~(\ref{e:sol1}), it is easy to see that the mean acceleration
imparted to the whole system by the radiation force is
$(\omega_r^2-\omega_e^2)x_{eq}=\omega_i^2 x_{eq}=\sigma_{es}F/(m_e+m_i)$,
as one could expect.  The
amplitude of the oscillation of the electrons relative to the ions is
$<\Delta x> \simeq x_{eq} \simeq {\cal L}$. Therefore, the average
electric field of the DL is $<E>\simeq 4\pi Z e n {\cal L}$. In the
non-relativistic limit, the velocity of the charged particles and the
DL electric field increase linearly with ${\cal L}$.

\begin{figure}
\epsscale{0.8}
\plotone{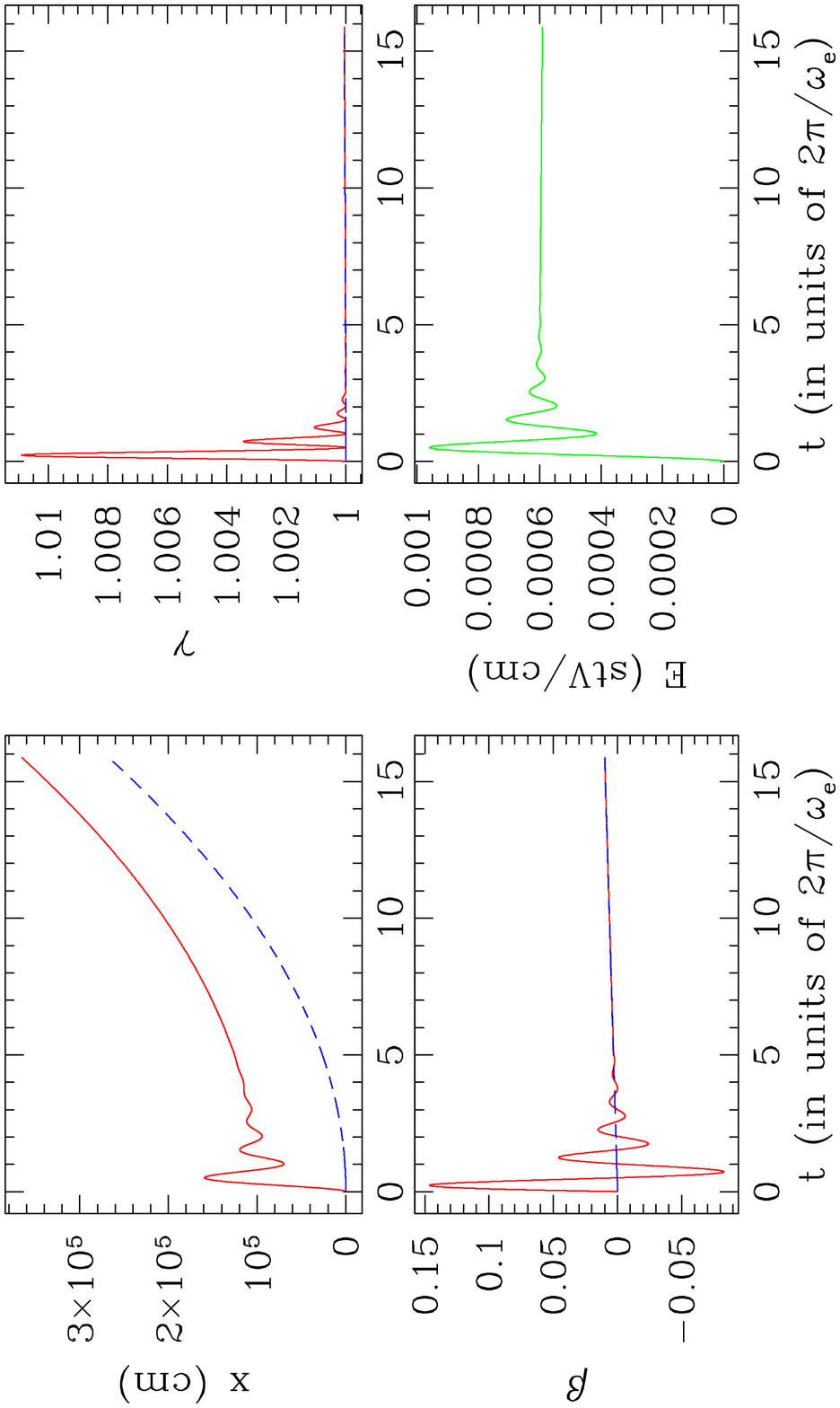}
\caption{Displacement $x$, velocity $\beta$ and
Lorentz factor $\gamma$ of electrons ({\it solid} line) and ions
({\it dashed} line) for ${\cal L} = 10^5$ cm and $\tau = 15.9
(2\pi/\omega_e)$ s. For $R=10^9$ cm, $n=1$ cm$^{-3}$ and $Z=1$
this corresponds to $L=1.6 \times 10^{36}$ erg s$^{-1}$ and
$\tau=1.8$ ms.  The induced DL electric field $E$ is also shown.
Here the energy of the monochromatic radiation spectrum
($\epsilon_0/m_ec^2=0.01$) is sufficiently low that the
Klein-Nishina correction to the scattering force is negligible.}
\label{fig3}
\end{figure}

\begin{figure}
\epsscale{0.8}
\plotone{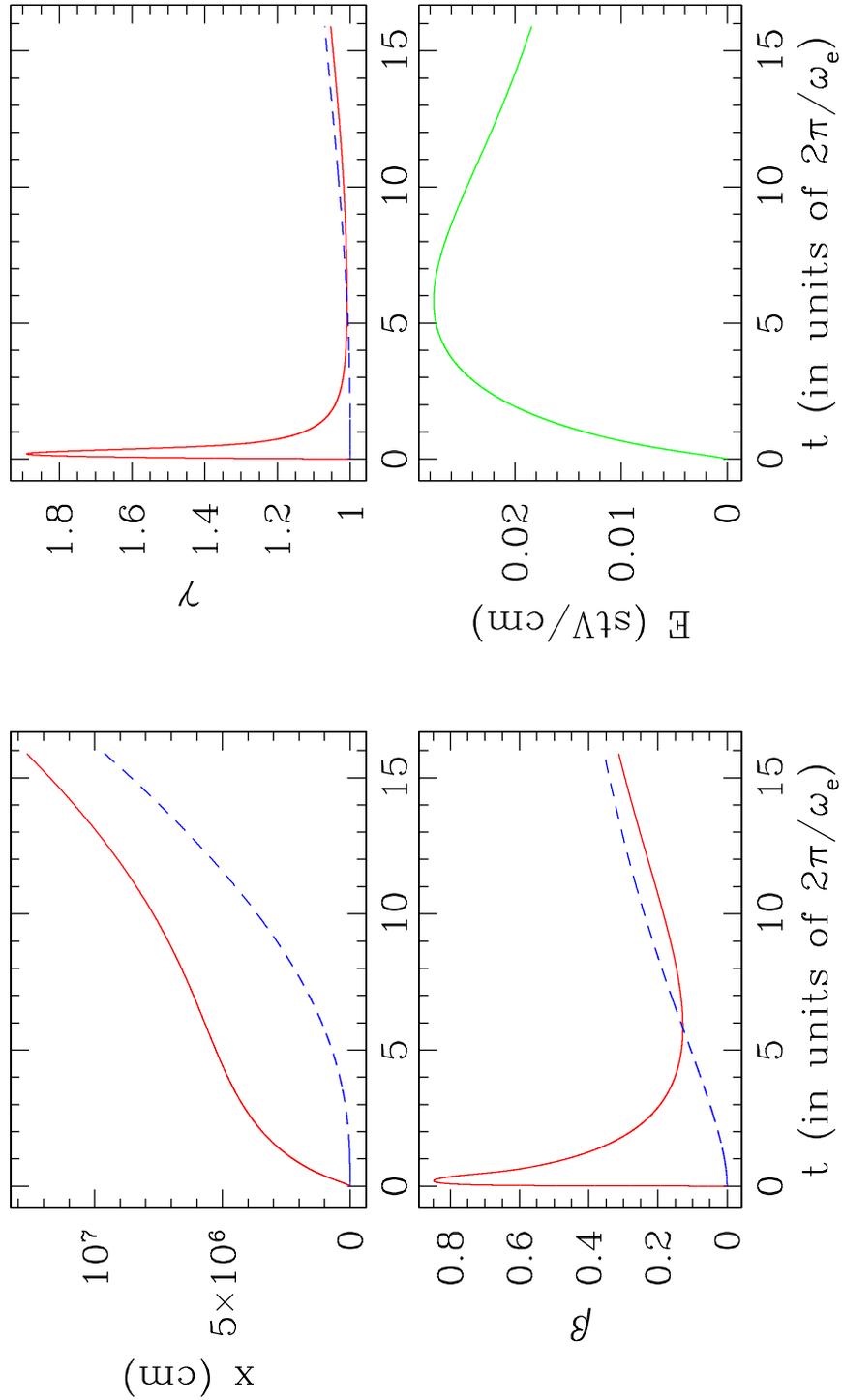}
\caption{Same as Fig.~\ref{fig3} for ${\cal L} = 6 \times 10^6$ cm
and $\tau = 15.9 (2\pi/\omega_e)$ s. For $R=10^9$ cm, $n=1$ cm$^{-3}$ and
$Z=1$ this corresponds to $L=9.7 \times 10^{42}$ erg s$^{-1}$
and $\tau=1.8$ ms.}
\label{fig4}
\end{figure}

\begin{figure}
\epsscale{0.8}
\plotone{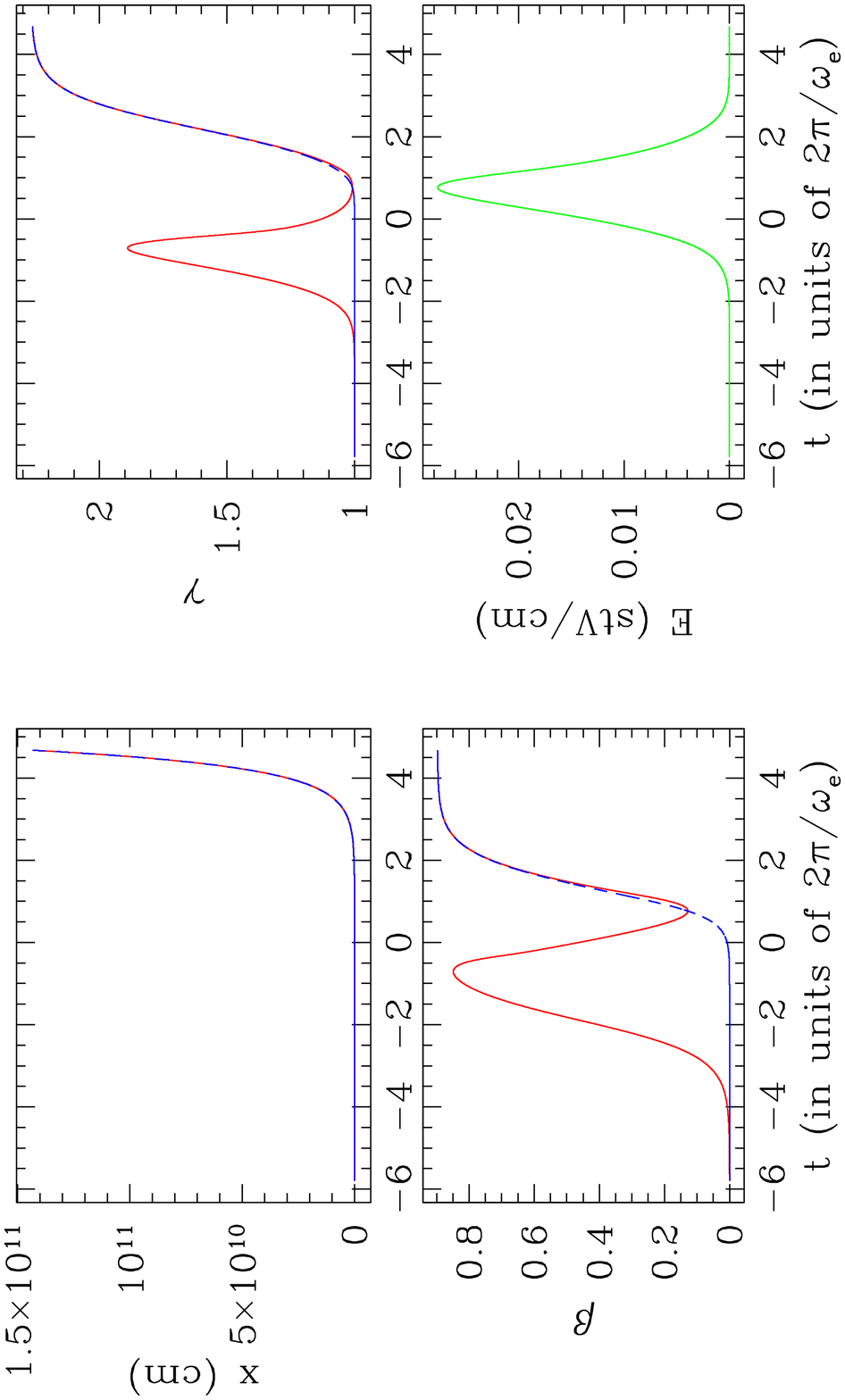}
\caption{Same as Fig.~\ref{fig3} for ${\cal L} = 6 \times 10^6$
cm and $\tau = 4.8 \times 10^4 (2\pi/\omega_e)$ s. For $R=10^9$
cm, $n=1$ cm$^{-3}$ and $Z=1$ this corresponds to $L=9.7
\times 10^{42}$ erg s$^{-1}$ and $\tau=5.4$ s. The scale on the
$x$-axis is logarithmic.} \label{fig4b}
\end{figure}

\begin{figure}
\epsscale{0.8}
\plotone{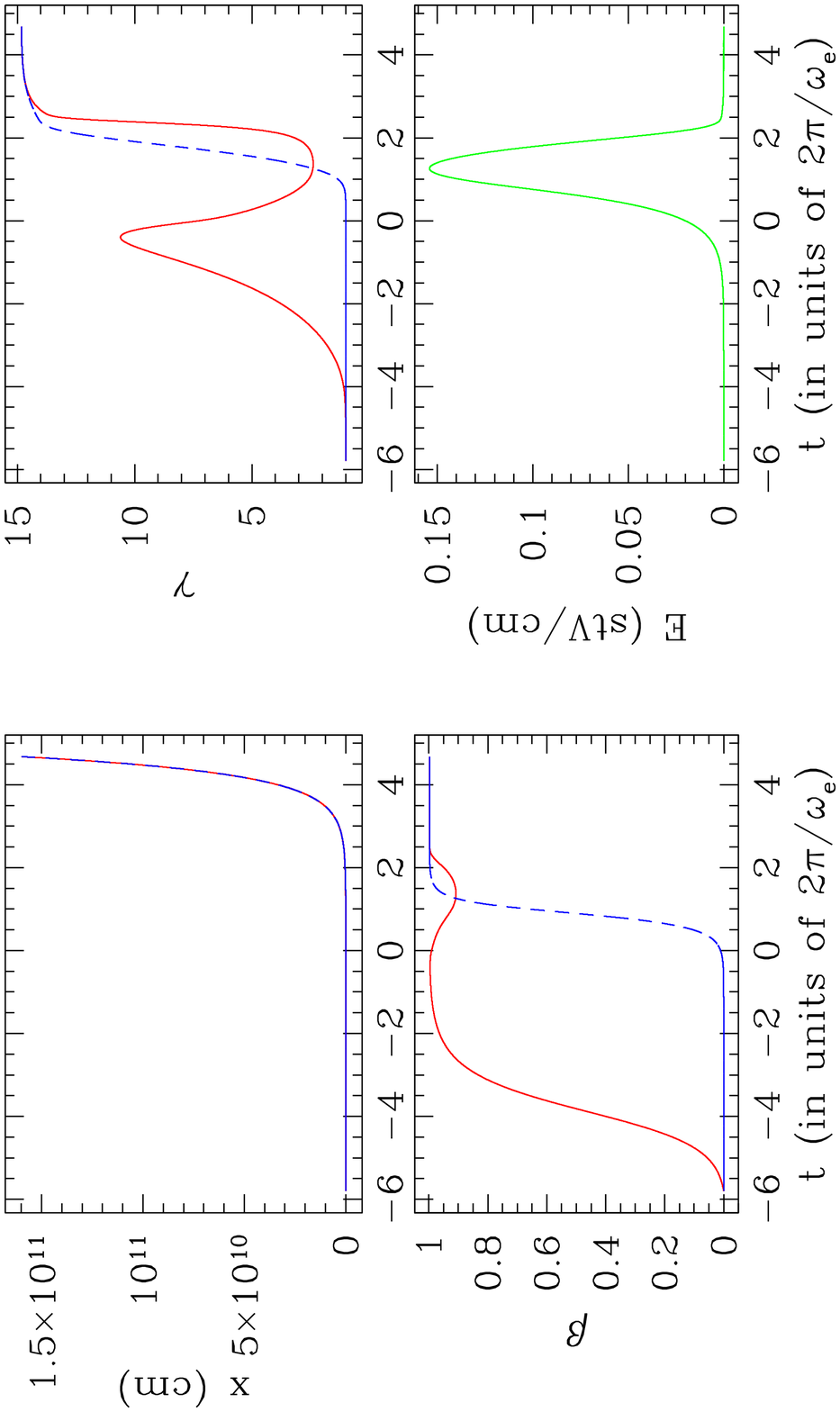}
\caption{Same as Fig.~\ref{fig3} for ${\cal L} = 6 \times 10^8$ cm
and $\tau = 4.8 \times 10^4 (2\pi/\omega_e)$ s. For $R=10^9$ cm, $n=1$
cm$^{-3}$ and $Z=1$ this corresponds to $L=9.7 \times 10^{44}$ erg
s$^{-1}$ and $\tau=5.4$ s. The scale on the $x$-axis is logarithmic.}
\label{fig5}
\end{figure}

Fig.~\ref{fig3} shows the numerical solution of the relativistic
equations (eqs.~[\ref{e:rel3}] and~[\ref{e:rel4}]) for a low
luminosity ($L=1.6 \times 10^{36}$ erg s$^{-1}$), short duration
($\tau=1.8$ ms) flash. The energy of the input monochromatic radiation
spectrum is $\epsilon_0/m_ec^2=0.01$, sufficiently low that the
Klein-Nishina correction to the scattering force is negligible.
During the early evolution, the essential properties of the
non-relativistic solution are recognizable. However, after a few
plasma time-scales, oscillations are significantly damped. In fact, in
the relativistic equations (e.g. eqs.~[\ref{e:rel1}]
and~[\ref{e:rel2}]) the radiative term contains a dependence on the
electron velocity $v_e$ that gives rise to a force formally similar to
a ``viscous'' force. However, this is not a physical drag exerted by
the radiation field onto the electrons because, by assumption, photons
are streaming radially. This effect is simply a consequence of the
Lorentz transformation for the radiation moments. In fact, the
radiative flux measured in the comoving frame varies with respect to
that measured in the system frame according to
equation~(\ref{e:flusso}). In radial streaming, the term
$v_e(U+P)$ becomes $2\beta_e F$.
This accounts for the decrease (or increase) in the flux measured in
the comoving frame caused by the electron motion and, ultimately,
originates from the finite velocity of propagation of photons. This
effect is already important at fairly modest electron speeds and, when
$v_e \geq 0.01-0.1 c$, the decrease of the radiative flux makes
radiative acceleration less efficient. The change in the sign of $v_e$
induced by the plasma oscillations causes a decrease/increase of the
radiative force (as seen in the comoving frame) that produces the same
effect of a viscous force, damping the oscillations of the electrons
on a few characteristic plasma time-scales (see Fig.~\ref{fig3})

\begin{figure}
\epsscale{0.8}
\plotone{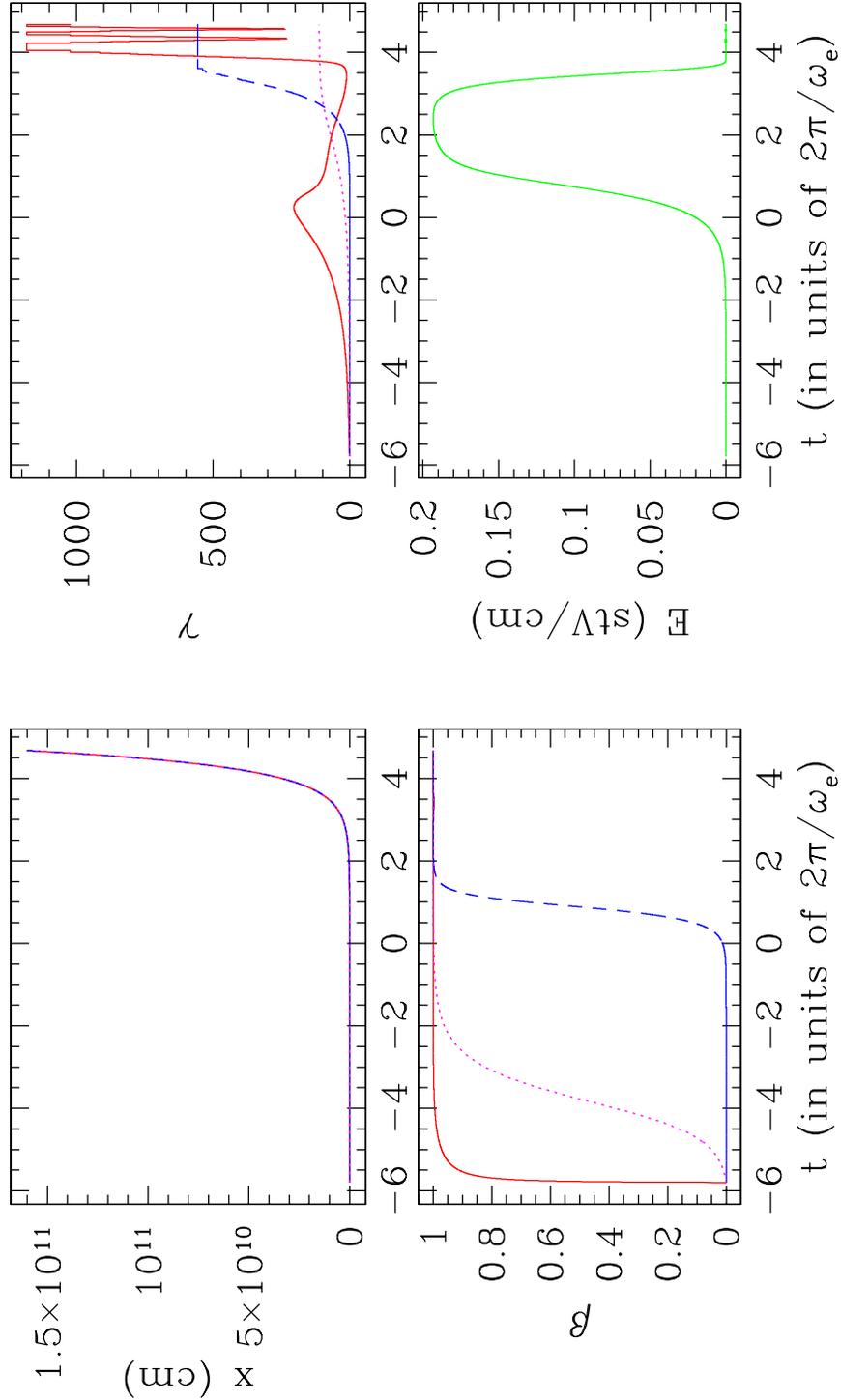}
\caption{Same as Fig.~\ref{fig3} for ${\cal L} = 10^{12}$ cm and
$\tau = 4.8 \times 10^4 (2\pi/\omega_e)$ s. For $R=10^9$ cm, $n=1$
cm$^{-3}$ and $Z=1$ this corresponds to $L=1.6 \times 10^{48}$ erg
s$^{-1}$ and $\tau=5.4$ s. The scale on the $x$-axis is
logarithmic. The {\it dotted} line represents the motion of a single
particle with total mass ($m_e+m_i$) and electron scattering cross
section $\sigma_T$.} \label{fig6}
\end{figure}

\begin{figure}
\epsscale{0.8}
\plotone{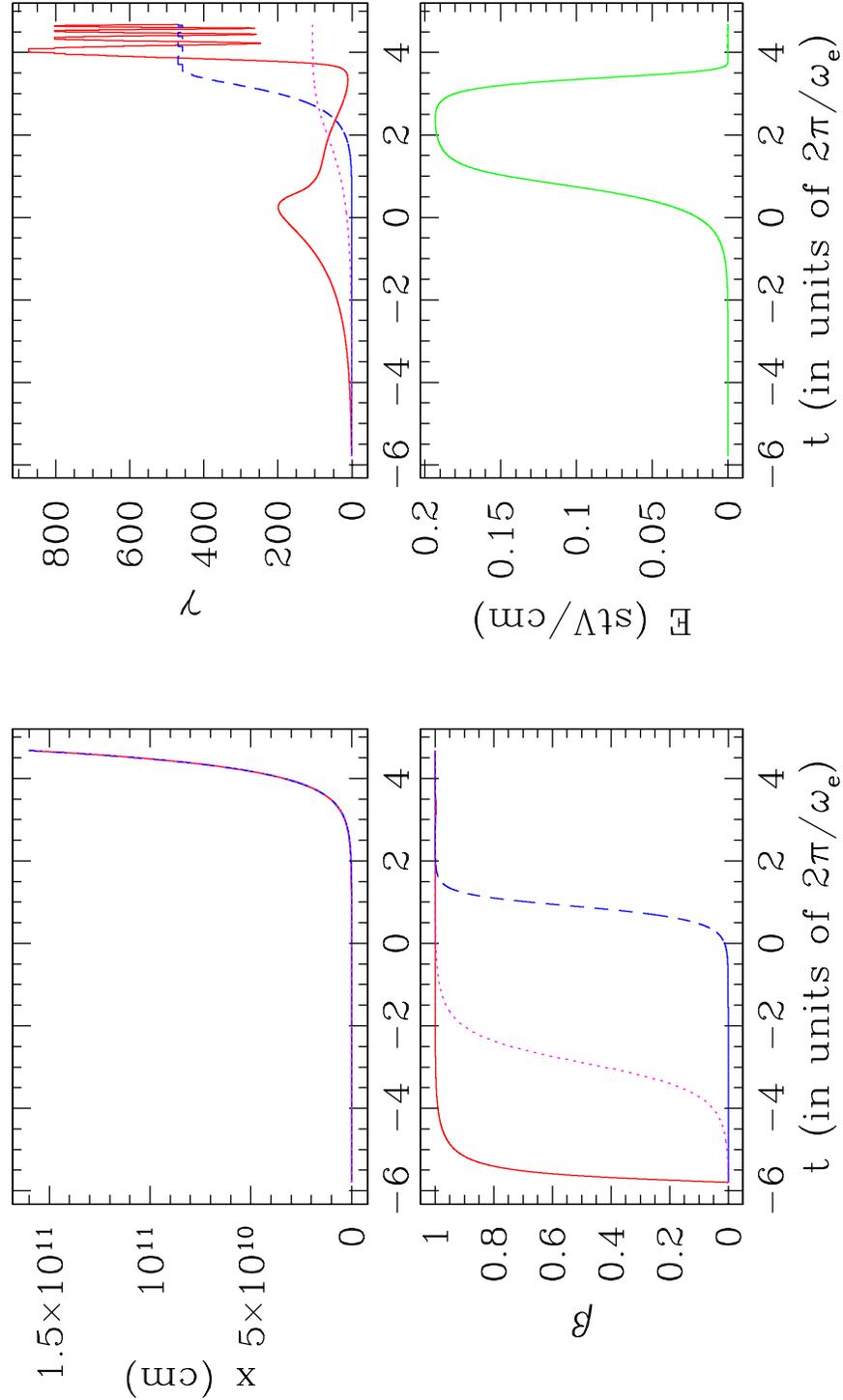}
\caption{Same as Fig.~\ref{fig3} for ${\cal L} = 10^{12}$ cm and
$\tau = 4.8 \times 10^4 (2\pi/\omega_e)$ s. For $R=10^9$ cm, $n=1$
cm$^{-3}$ and $Z=1$ this corresponds to $L=1.6 \times 10^{48}$ erg
s$^{-1}$ and $\tau=5.4$ s.  Here the energy of the monochromatic
radiation spectrum is assumed to be $\epsilon_0/m_ec^2=10$.  The scale
on the $x$-axis is logarithmic. The {\it dotted} line represents the
motion of a single particle with total mass ($m_e+m_i$).} \label{fig8}
\end{figure}

For ${\cal L} > 10^6$~cm, the system approaches the truly
relativistic regime. As shown in Fig.~\ref{fig4}, during the
initial transient phase, the radiative force is so strong that
electrons are accelerated rapidly to relativistic speeds while
the ions lag behind owing to their larger inertia. This produces
an electric restoring force that accelerates the ions for all the
duration of the burst. Oscillations are completely damped. For a
short duration flash, the final speeds and energies of the
charged particles are rather modest. However, as shown in
Fig.~\ref{fig4b}, for a long duration burst ($\tau=5.4$ s), the
asymptotic speeds approach the velocity of light. This is more
clearly shown in Figs.~\ref{fig5} and~\ref{fig6}, where the
initial transient is so rapid that $v_e \simeq c$ in less than a
characteristic plasma time-scale and the ions are later
effectively accelerated by the strong radiation-induced DL
electric field up to $\gamma \sim 500$.
When the ions catch up with the electrons at $t\sim 2000 (2\pi/\omega_e)$,
the electric restoring force becomes negligible
and the electrons are rapidly re-accelerated by the radiation
field until their velocity becomes equal to that of the ions. In
this asymptotic regime, the DL has already moved significantly
from the initial position ($r > 10^{10}$ cm) and the Lorentz
factor has become so large that the radiative flux in the
comoving frame becomes negligible.
Radiative acceleration is no longer very efficient and electrons and
ions are almost freely streaming, with the electrons oscillating
around the ions (see Fig.~\ref{fig6}). The asymptotic large amplitude
oscillations achieved by the electrons may give rise to strong
internal dissipation, thereby increasing the plasma temperature.

Models with ${\cal L}\ga 10^{11}$ cm suffer, however, from two major
limitations. In fact, assuming $R\sim 10^9$~cm, the luminosity exceeds
$10^{47}$~erg s$^{-1}$ and the effective temperature of the radiation
is $\ga 10^9$~K for a typical source size $R_s \approx 10^7$~cm. The mean
photon energy is then sufficiently large that the Klein-Nishina
correction to the radiative force (see eq.[\ref{e:fradrels}]) starts
to become significant. This is a fortiori true if the primary photon
spectrum has a significant high-energy component. An example of the
particles dynamics in this regime is illustrated in Fig.~\ref{fig8}
for monochromatic input photons with $\epsilon_0/m_ec^2=10$. As
expected, the main effect of the reduction in the effective
cross-section is to lower the terminal Lorentz factor of both ions and
electrons, but this does not change the essential features of the
acceleration process discussed above. The second issue concerns the
possibility that the wind may become pair-loaded as a consequence of
$\gamma\gamma$ interactions among primary photons or between primary
and once-scattered photons. This effect is bound to
increase dramatically the efficiency of radiative acceleration because
it reduces the inertia of the plasma, as discussed in detail by
\citet{THOMPSON}. Charge-separation in a pair-dominated medium is
outside the scope of the present investigation, so in the following we
restrict to cases with photon energies well below the threshold for
pair production. For thermal radiation with effective temperature
$\la 5 \times 10^8$ K, the fraction of photons with energy two times
larger than the threshold energy ($\sim 7$ MeV, at which the cross section
for pair-production peaks) is
$f \approx 10^{-54}$. The luminosity from a source of characteristic size
$\sim 10^7$ cm is $L \la 5 \times 10^{45}$ erg s$^{-1}$, corresponding to
${\cal L} \la 3 \times 10^9$ cm for $R=10^9$ cm.
The optical depth for $\gamma\gamma$ interaction between two
primary photons close to the source is 
$\tau_{\gamma\gamma} \approx n_{\gamma\gamma} \sigma_{\gamma\gamma} R_s$,
where $\sigma_{\gamma\gamma} \approx \sigma_T$ for large scattering
angles and $n_{\gamma\gamma}\sim (L/\epsilon) \Delta t/R_s^3$ for a
burst of duration $\Delta t$. Although $\tau_{\gamma\gamma}$ would be lower
for beamed or anisotropic emission, close to an isotropic source it is
much larger than unity and all photons above threshold can produce
$e^{+}e^{-}$ pairs. However, for $T \la 5 \times 10^8$ K, there are
no photons with energy two times larger than the threshold energy
($N_1 \approx f (L\Delta t/\epsilon) \ll 1$) and hence the radiation
field is pair-free. As for the photons scattered in an off-radial
direction by the plasma electrons at radius $R$,
their optical depth to primary photons is $\approx p \tau_{\gamma\gamma}$,
where $p\approx n\sigma_T \Delta x < 1$ is the scattering probability.
The value of $p \tau_{\gamma\gamma}$ can be larger or smaller than unity,
depending on the plasma density. But, even if $p \tau_{\gamma\gamma} > 1$,
for $T \la 5 \times 10^8$ K no additional pair-loading is induced because
there are no primary photons above threshold and once-scattered photons
did not gain appreciable energy in the scattering.
Therefore, for ${\cal L} < 10^{10}$ cm the dynamics of the plasma
shells is not influenced by pair-loading.

For these values of the parameter ${\cal L}$, the probability that
once-scattered photons collide with another electron is actually
larger than undergoing $\gamma\gamma$ interaction. If at the time of
the second scattering event the electron has already acquired
relativistic velocity, in the electron rest frame the energy of the
scattered photon is boosted by a factor $\gamma$. So, if $p$ is the
probability of the first scattering, the probability that a photon
undergoes two scatterings is $p^2/\gamma$ (owing to the decrease in
the Klein-Nishina cross section). Then, the number of twice scattered
photons for a $\Delta t = 1$ s burst is
$\approx (L\Delta t/\epsilon) p^2/\gamma \sim 10^{50} p^2 \, (L/10^{46} {\rm erg \ s}^{-1})$.
In order not to influence significantly the dynamics of the plasma
shells, this number must be smaller than $\approx nR^2\Delta x$, which
implies $n \la 10^{11}$ cm$^{-3}$ for $\Delta x \approx 10^7$ cm.
For plasma densities in excess of
$n\sim 10^{11}$ cm$^{-3}$, the dynamics of the plasma shells starts to
be significantly influenced by Compton up-scattering of off-radial
photons (Compton drag). Twice-scattered photons are boosted to very
large energies. Energy conservation limits their maximum energy to
$\gamma m_e c^2 \la 50$ MeV, corresponding to a the total luminosity
$\approx 10^{46} p^2$ erg s$^{-1}$.

Compton drag caused by off-radial primary photons may also play
an important role in the vicinity of the source. In order to understand
qualitatively this effect, we solved the equations assuming a
slight deviation of the radiation field from the perfect radial
case (of the order of $R_s^2/R^2 \sim 10^{-4}$). The solution is
shown in Figure~\ref{fig9} where, in order to avoid the regime of
pair-loading, ${\cal L} \simeq 3 \times 10^9$~cm. Compton drag
decreases the electron Lorentz factor attainable during the
electrostatic ion acceleration stage. Therefore, once the ions
have been accelerated to relativistic energies by the
radiation-induced DL electric field, they overtake the electrons
and the DL reverses. At this stage the electrons are
re-accelerated at the expense of the electrostatic potential
energy of the DL and, ultimately, of the kinetic energy of the
ions. The asymptotic Lorentz factors of ions and electrons are
about 10 times smaller than those achieved for a perfectly radial
radiation field (point source). Placing the plasma shell closer
to the central source would end up in a sufficient deviation from
radial streaming to cause too a large decrease of the overall
efficiency of the acceleration mechanism.

We note that, even if the plasma is initially cold, random internal
energy will be imparted to the electrons by the incident hard-X and
gamma-rays, because direct energy and momentum transfer results when
outward-directed photons are removed from nearly radially streaming
radiation field and scattered away. In principle, if a sufficient
amount of internal energy is transferred to the electron plasma, it
can be effectively converted into bulk motion through the so-called
``Compton rocket'' (\citealt{ODELL}; \citealt{ODELLb};
\citealt{CHENG}). In this way, the energy lost to heat may be
compensated by that gained by the electrons thanks to the additional
Compton thrust. Although the study of this regime is beyond the scope
of the present investigation, the outcome of the electron acceleration
process with the additional Compton thrust may not be qualitatively
different with respect to what found here.

The radiation-induced electrostatic acceleration of the plasma shells
is effective if they are quite spatially concentrated and overdense
with respect to the surrounding medium. In fact, if the shells are
spatially extended, the escaping radiation would accelerate the plasma
located at larger and larger radii at progressively lower speeds.
The innermost, high velocity DLs would rapidly overtake the outer, low
velocity ones, producing shocks and dissipation of kinetic energy into
heat.

It is noteworthy that, for intense radiation fluxes (${\cal
L}>10^8-10^9$~cm), the asymptotic energies achieved by both ions and
electrons are larger by a factor 2--4 (depending on the burst
duration, see Fig.\ref{fig9}) with respect to what one
could naively expect assuming that the electron-ion system was
rigidly coupled with a total mass ($m_e+m_i$) and electron
scattering cross section $\sigma_T$ (see e.g. \citealt{GUREVICH}).
This is caused by the fact that the radiative
force is not conservative and hence the dynamical evolution of
the system is of fundamental importance in determining the
asymptotic energies. Thanks to the prompt response of the
electrons, the work done by the radiation field is accumulated in
the DL and is later effectively converted into kinetic energy of
the ions through the acceleration produced by the DL electric
field. As shown in Fig.~\ref{fig9}, this process is significantly
more effective than radiatively accelerating a rigidly coupled
electron-ion system. In fact, ions become relativistic at $t_1
\simeq 30 (2\pi/\omega_e)$ and are electrostatically accelerated
to large Lorentz factors up to $t_2 \simeq 300 (2\pi/\omega_e)$,
when they catch up with the electrons.  In the interval
$t_1\la t \la t_2$, the average velocity of the electrons is smaller
then that of a rigidly coupled system and hence the radiative
force is more effective. The net work done by the radiation field
is 2--4 times larger than it would otherwise be and is stored in
the DL as electrostatic potential energy. The maximum electric
field and energy can be estimated assuming that, initially
($0\leq t \leq t_1$), the electrons travel at the speed of light
while the motion of the ions is approximately non-relativistic.
Then, $x_e \simeq ct$ and $d^2x_i/d(\omega_e t)^2 \simeq
(m_e/m_i) x_e = (m_e/m_i) ct$. Integrating this equation with
$x_i(0)=0$ and $dx_i/dt|_0=0$, we can calculate the time $t_1$ at
which $v_i\simeq c$ and the separation $\Delta x_{max}$ at
$t_1$.  It is: $t_1= (2m_i/m_e)^{1/2} (1/\omega_e) \simeq 10
(2\pi/\omega_e)$ and $\Delta x_{max}=(2\sqrt{2}/3)
c/\omega_i\simeq 2 \times 10^7$ cm. Substituting these
expressions in equation~(\ref{e:campoel}), we obtain
\begin{equation}
E_{max} = \frac{8\pi\sqrt{2}}{3} \frac{Z e n c}{\omega_i}
\simeq 0.13 \left(\frac{n}{1 \ {\rm cm}^{-3}}\right)^{1/2}
{\rm stV \ cm^{-1}} \, ,
\label{e:emax}
\end{equation}
in quite good agreement with the maximum electric field reported
in Fig.~\ref{fig9}. The distance traveled by the ions during the
relativistic acceleration phase ($t_1 \leq t \leq t_2$) is
$\simeq c (t_2-t_1)$ and then their asymptotic energy is
approximately given by $W_{max} \simeq e E_{max} c (t_2-t_1)
\simeq 10^{11}$ eV.

Despite the large asymptotic energies achieved by ions and electrons,
the overall efficiency of this process is very low. The final energy
acquired by an ion is
$E_{tot} \sim Z m_p c^2 \gamma_\infty \sim 5 \times 10^{-2} Z \, (\gamma_\infty/100)$ erg. Comparing this energy with the total work done by the radiation
field on $Z$ electrons,
$W_{rad} \sim Z L \tau \sigma_T/4\pi R^2 \sim 5 \times 10^2 Z \, (L/10^{46} \, {\rm erg})$ erg, we find that the efficiency is not larger than $\sim 10^{-4} \, (\gamma_\infty/100)$.

\section{Conclusions}

We have shown that radiation-induced electric fields can increase
significantly the terminal energy of a ion-electron plasma accelerated
by a transient radiation source. The asymptotic Lorentz factors are a
factor 2--4 larger than those attainable by radiative acceleration
alone. The effect is maximum when both radiation drag and
$\gamma\gamma$ interactions are negligible, i.e. when the central
source is distant ($R_s^2/R^2 \la 10^{-4}$) or beamed, the plasma
density is low ($n \la 10^{11}$ cm$^{-3}$) and the average photon
energy is well below the threshold for pair production ($T\la 5\times
10^8$ K). For $R\simeq 10^9$ cm, these limitations imply $L\la 5\times
10^{45}$ erg s$^{-1}$, or ${\cal L} \la 3 \times 10^9$~cm.
Decreasing/increasing the initial position of the plasma shell by an
order of magnitude results in a corresponding decrement/increment of
the maximum allowed luminosity ($L \la 5\times 10^{43}$ erg s$^{-1}$
and $L \la 5\times 10^{47}$ erg s$^{-1}$, respectively). However,
because ${\cal L}$ remains the same and, for a fixed duration burst,
is proportional to the total work done by the radiation field on the
ion-electron assembly, the dynamics and terminal energy are similar in
all these cases.

\begin{figure}
\epsscale{0.8}
\plotone{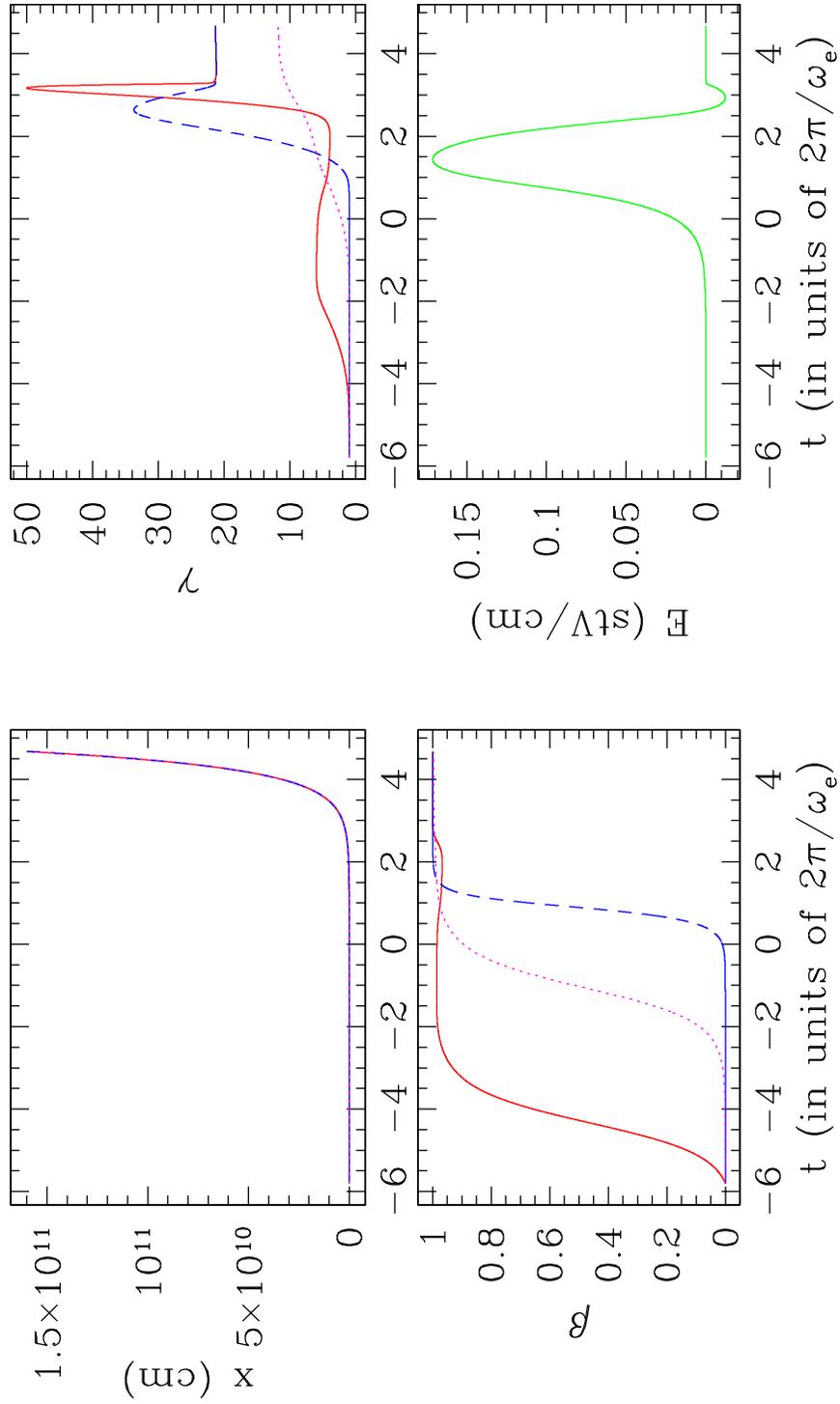}
\caption{Same as Fig.~\ref{fig8} for ${\cal L} = 2.8 \times 10^9$
cm and $\tau = 4.8 \times 10^4 (2\pi/\omega_e)$ s. For $R=10^9$ cm,
$n=1$ cm$^{-3}$ and $Z=1$ this corresponds to $L=4.5 \times 10^{45}$ erg
s$^{-1}$ and $\tau=5.4$ s.  Here the anisotropy of the radiation field
has been accounted for (see text for details).}
\label{fig9}
\end{figure}

Although direct astrophysical applications are outside the scope of
the present investigation, it is quite interesting to note that giant
flares from soft gamma repeaters (SGRs) are typically within the
parameter range considered here. The total energy emitted in two giant
flares observed from SGR 0525-66 and SGR 1900+14 was $\approx
10^{44}-10^{45}\,{\rm erg}$ on a timescale $\approx 1$~s, with an
implied luminosity of about the same order (see e.g. \citealt{HURLEY}
for a review). In the picture proposed by \citet{THOMPSONDUNCAN} the
giant flare is produced by a large-scale instability of the neutron
star magnetic field which leads to the formation of an expanding pair
fireball. The radiative acceleration process presented in this paper
has no direct connection with the dynamics of the baryon-sheathed pair
fireball, whose characteristic size is $\approx 10^6-10^7$~cm.
However, it is possible that the outflowing radiation front impinges
on a low-density ordinary plasma at much larger distances, $\approx
10^9$~cm, accelerating protons and electrons to relativistic
velocities as illustrated e.g. in Fig.~\ref{fig9}.

\acknowledgments
This work has been partially supported by the Italian Ministry for
Education, University and Research (MIUR) under grants
COFIN-2000-MM02C71842 and COFIN-2002-027145. We thank an anonymous
referee for helpful comments.

\clearpage

\clearpage

\clearpage

\clearpage

\clearpage

\clearpage

\clearpage

\end{document}